\documentclass[prb,aps,amsmath]{revtex4} 
\usepackage{amssymb}
\usepackage{latexsym}
\usepackage{graphicx}
\usepackage{hyperref}
\usepackage[english]{babel}
\hypersetup{linktocpage,,colorlinks=true}
\usepackage{color}
\usepackage{mathtools}
\usepackage{epsfig}
\usepackage{subfigure}
\usepackage{dcolumn}
\usepackage{amsmath}
\usepackage{amsfonts}
\usepackage{bm}
\begin{document}
\title{A Theoretical Study of Doping Evolution of Phonons in High-Temperature Cuprate Superconductors}
\author{ Saheli Sarkar}
\affiliation{Division of Condensed Matter Physics and Materials Science, Brookhaven National Laboratory, Upton, NY 11973-5000, USA}
 \date{\today } 
  \begin{abstract} 
 Hole-doped high-temperature copper oxide-based superconductors (cuprates) exhibit complex phase diagrams where electronic orders like a charge density wave (CDW) and superconductivity (SC) appear at low temperatures. The origins of these electronic orders are still open questions due to their complex interplay and correlated nature. These electronic orders can modify the phonons in the system, which has also been experimentally found in several cuprates as a softening in the phonon frequency at the CDW vector.
Recent experiments have revealed that the softening in phonons in cuprates due to CDW shows intriguing behavior with increasing hole doping. Hole doping can also change the underlying Fermi surface. Therefore, it is an interesting question whether the doping-induced change in the Fermi surface can affect the softening of phonons, which in turn can reveal the nature of the electronic orders present in the system. In this work, we investigate this question by studying the  softening of phonons in the presence of CDW and SC within a perturbative approach developed in an earlier work. We compare the results obtained within the working model to some experiments.
 \end{abstract}

\maketitle

\section{Introduction}

High-temperature copper oxide-based superconductors (cuprates) are paradigmatic examples of a complex interplay between several orders like a quasi-two-dimensional incommensurate charge density wave (CDW) \cite{ref-HoffmanSc2002,ref-GhirSc2002} and superconductivity (SC) \cite{ref-TsueiRMP200}, especially in the under-doped regime of hole-doped cuprates~\cite{ref-FradRMP2015}. 
Besides~these orders, in~the under-doped regime, there exist a `pseudogap' phase~\cite{ref-AlloulPRL1989,ref-RennerPRL1998,ref-InoPRL1998} and a variety of other complex orders~\mbox{\cite{ref-EmeryPNAS1999,ref-FauquePRL2006,ref-HeScie2011,ref-SarkarPRB2019}}. In~addition to the observation of the CDW order in the under-doped regime, recent X-ray scattering experiments have observed CDW in over-doped samples of LSCO~\cite{ref-WenNat2019,ref-vonNPJ2023}, even going well beyond the pseudogap region~\cite{ref-LiPRL2023}, and in over-doped BSCCO~\cite{ref-PengNat2018}. Furthermore, recent resonant X-ray scattering experiments have observed charge density modulations with correlation length much shorter compared to the CDW which are generally known as charge density fluctuations (CDFs) \cite{ref-ArpiaNatCom2023}. These CDFs have a dynamical nature due to finite energy and are present in an extremely large doping window starting from a highly under-doped regime and going to a highly over-doped regime~\cite{ref-ArpiaScience2019,ref-ArpiaNatCom2023}. Additionally, these dynamical CDFs persist up to a very high temperature, above~the onset of CDW and even above the pseudogap temperature T$_{PG}$.

A change in doping not only tunes the electronic orders, but~also changes the electronic structure and the associated Fermi surface. Angle-resolved photoemission spectroscopy (ARPES) experiments~\cite{ref-Fujimori1998,ref-YoshidaPRB2006,ref-ZhongPNAS2023} measured the evolution of the Fermi surface of La$_{2-x}$Sr$_x$CuO$_4$ (LSCO) for a broad range of hole doping starting from the under-doped regime and going to the over-doped regime. Another ARPES experiment~\cite{ref-KAMPRB2006} measured the hole doping evolution of the Fermi surface of Bi$_{2}$Sr$_2$CaCu$_2$O$_{8 +\delta}$ (BSCCO). For~both the cases of LSCO and BSCCO, it is observed that the geometry of the Fermi surface changes from an open `hole-like' to a closed `electron-like' geometry. This type of topological transition of the Fermi surface is known as the Lifshitz~transition. 

Owing to such a complex phase diagram and Fermi surface evolution, the~nature of the interactions which induce the electronic orders still remains an open question. A~possible direction for disentangling various interactions is to study the collective modes associated with the electronic orders~\cite{ref-LoretNat2019}. It is well known that CDW can couple to the phonons and in metallic systems give rise to the softening of the phonon frequency (we refer to it as phonon softening) at the wave vector associated with the CDW ($\bm{Q}$) and below the CDW transition temperature. This is known as the `Kohn anomaly'~\cite{ref-WollPR1962,ref-RenkerPRL1973}. Interestingly, experiments have found signatures of phonon softening around $\bm{Q}$ but below the superconducting critical temperature (T$_c$) in several under-doped cuprates like YBa$_2$Cu$_3$O$_{7-x}$ (YBCO), Bi$_{2}$Sr$_2$CaCu$_2$O$_{8 +\delta}$, La$_{2-x}$Ba$_x$CuO$_4$ (LBCO) and La$_{2-x}$Sr$_x$CuO$_4$ \cite{ref-TaconNAT2014,ref-LeeNAT2021,ref-ReznikNat2006,ref-McQuPRL1999,ref-LinPRL2020}. In~under-doped La-based 214 compounds~\cite{ref-LinPRL2020} and BSCCO~\cite{ref-LeeNAT2021,ref-WangQScAd2021}, however, the~phonon softening does not completely vanish at T$_c$. Besides~the temperature evolution, the~phonon softening in different cuprates also exhibits distinct features with a change in doping. Resonant inelastic X-ray scattering (RIXS) and neutron scattering experiments investigated phonons for different doping levels in Bi$_{2}$Sr$_2$CaCu$_2$O$_{8 +\delta}$ \cite{ref-LuPRB2022}, La$_{2-x}$Sr$_x$CuO$_4$ \cite{ref-LinPRL2020} and La-based 214~compounds~\cite{ref-ReznikNat2006}, respectively. Remarkably, for~doped BSCCO, the~RIXS experiment~\cite{ref-LuPRB2022} found a gradual suppression of phonon softening with an increase in doping in a broad window of an under-doping to an over-doping regime. In~contrast to BSCCO, however, in~doped LSCO~\cite{ref-LinPRL2020}, it was found that phonon softening related to charge density modulations remained unchanged in a broad window of hole doping despite a continuous change in the Fermi surface~\cite{ref-YoshidaPRB2006}, before~vanishing discontinuously at a critical~doping.

To explain the phonon softening below T$_c$, in~an earlier work~\cite{ref-SarkarPRR2021}, we investigated collective modes of phonons when they couple to both CDW and SC orders within a perturbative approach. We showed, for prototypical under-doped cuprates' band structure, that a~complex interplay between the CDW and SC orders and their fluctuations~\cite{ref-PepinARC2020} can capture the temperature evolution of the phonon softening, especially the anomalous phonon softening below T$_c$. 
However, the~evolution of the phonon softening with a change in doping was not addressed. It is especially interesting to see if the change in the Fermi surface due to doping can be captured through collective phonon modes, which can also elucidate the nature of the electronic orders present in the~system.

Motivated by these, in~this work we study, within our theoretical model developed in~\cite{ref-SarkarPRR2021}, the~evolution of phonon softening for cuprates with a change in the geometry of the Fermi surface. In~particular, we focus on two model cuprate bands replicating doped LSCO and BSCCO whose Fermi surfaces pass through a topological Lifshitz transition, where a hole-like Fermi surface transforms to an electron-like Fermi surface. We study the fate of the phonon softening approaching this transition for both of the model~cuprate bands. 

We organize this paper in the following manner. In~Section~\ref{sss222}, we briefly discuss the theoretical model, where the effects of CDW and SC orders on phonons are investigated. In~Section~\ref{sss333}, we study the doping evolution of the phonon softening for model Fermi surfaces representing doped LSCO and BSCCO systems. In~Section~\ref{sss444}, we present a discussion of our work and compare with experimental observations. In~Section~\ref{sss555}, we present a summary of the~work.

\section{Theoretical~Model}\label{sss222}

In this section, we describe the theoretical model of a coupled electron--phonon system in the presence of the CDW and SC orders. The~total Hamiltonian~\cite{ref-SarkarPRR2021} is given by $H_{tot} = H_{e} + H_{ph} + H_{e-ph}$, with
\vspace{-6pt}
\begin{align}\label{eqn:Ham}
H_{e} & = \sum_{\bm{k},\sigma} \xi_{\bm{k}}c_{\bm{k},\sigma}^{\dagger}c_{\bm{k},\sigma} +
 \sum_{\bm{k},\sigma}(\chi_{\bm{k}}c_{\bm{k}+\bm{Q},\sigma}^{\dagger}c_{\bm{k},\sigma} + h.c.) +\sum_{\bm{k}}(\Delta_{\bm{k}}c_{\bm{k},\uparrow}^{\dagger}c_{-\bm{k},\downarrow}^{\dagger} + h.c.),\\
\nonumber
H_{ph} &= \sum_{\bm{q}}\omega_{\bm{q}}b_{\bm{q}}^{\dagger}b_{\bm{q}},\\
\nonumber
H_{e-ph} &= (1/\sqrt{N})\sum_{{\bm{q}}}\sum_{\bm{k},\sigma}g(\bm{k},\bm{q})c_{\bm{k}+\bm{q},\sigma}^{\dagger}c_{\bm{k},\sigma}(b_{-\bm{q}}^{\dagger}+ b_{\bm{q}}).
\end{align}

In Equation \eqref{eqn:Ham}, $H_e$ depicts the electronic Hamiltonian in the presence of CDW and SC orders at the mean-field level, where  $c_{\bm{k}}$ and $c_{\bm{k}}^{\dagger}$ are the electronic annihilation and creation operators. $\xi_k$,  $\chi_k$ and $\Delta_k$ are the tight-binding single-particle dispersion, CDW gap and SC gap, respectively. The~Hamiltonian for the phonons is given by $H_{ph}$, where $b_{\bm{q}}$ and $b_{\bm{q}}^{\dagger}$ are the phonon annihilation and creation operators and $\omega_{\bm{q}}$ is the bare phonon frequency. The~interaction between electrons and phonons is given by the Hamiltonian $H_{e-ph}$, where $g(\bm{k},\bm{q})$ is the electron--phonon (e-ph) coupling element and N is the number of lattice sites. In~our model, we consider $g(\bm{k},\bm{q}) =g$ a~constant number. We construct the matrix $H_e$ in the extended Nambu basis $\Psi^{\dagger}_{\bm{k}}  = \left(c^{\dagger}_{\bm{k},\uparrow},c_{-\bm{k},\downarrow},c^{\dagger}_{\bm{k}+\bm{Q},\uparrow},c_{-\bm{k}-\bm{Q},\downarrow}\right)$, where $\bm{Q}$ is the CDW wave vector. The~Green's function corresponding to $H_{e}$ is given by $\hat G^{-1}(i\omega_{n},\bm{k}) = (i \omega_{n} -\hat H_{e} )$ and has a matrix form as follows:
\begin{align}\label{eq_greenmat}
\tiny
G^{-1}&=
\begin{pmatrix}
i\omega_{n}-\xi_{\bm{k}} & -\Delta_{\bm{k}} & -\chi_{\bm{k}} &0 \\
-\Delta^{*}_{\bm{k}} & i\omega_{n}+\xi_{\bm{k}} & 0 &\chi_{\bm{k}}\\
-\chi^{*}_{\bm{k}} &0 & i\omega_{n} - \xi_{\bm{k}+\bm{Q}} & -\Delta_{\bm{k}+\bm{Q}}\\
0 & \chi^{*}_{\bm{k}} &-\Delta^{*}_{\bm{k}+\bm{Q}} & i\omega_{n}+\xi_{\bm{k}+\bm{Q}}
\end{pmatrix},
\end{align}
where $\omega_{n}$ is the Matsubara~frequency.

Now, we analyze the change in the phonon spectrum due to the presence of both CDW and superconducting order. Because~of the presence of the CDW with wave vector Q, the~new phonon propagator can be written in terms of a matrix Green's function given by $D_{m,n}(\bm{q},\tau) = -\langle \mathcal{T} \phi_{\bm{q}+m\bm{Q}}(\tau)\phi^{\dagger}_{\bm{q}+n\bm{Q}}(0)\rangle$ \cite{ref-PALeeSSC1993}, where $\mathcal{T}$ is the time-ordering operator and $\phi_{\bm{q}}$ is given by $b_{\bm{q}}^\dagger + b_{-\bm{q}}$ with $m,n = \pm$. We also note that $D_{++} = D_{--}:=D_1(z,\bm{q})$ and $D_{+-} = D_{-+}:=D_2(z,\bm{q})$.
The modified propagators can be calculated by treating the electron--phonon interaction perturbatively, as~long as g is small. For~a single-dispersion-less phonon mode,
 an~estimate for g can be obtained by using the dimensionless coupling constant $\lambda \sim \frac{g^2 N(E_F)}{\nu_0}$, where $N(E_F)$ is the density of states at the {\color{black}Fermi} energy and $\nu_0$ is the bare phonon frequency. As~typically $\frac{\nu_0}{N(E_F)} \ll 1$, the~approximation remains valid even for large values of $\lambda$.  Following the results obtained in the paper~\cite{ref-SarkarPRR2021}, the~self-energy corrections [$\Sigma_{1,2,3,4}(z,\bm{q})$] for phonon propagators due to the presence of both CDW and SC orders are
\begin{align}\label{eq:suple_pi}
\Sigma_{1}(\omega,\bm{q}) & = \frac{g^{2}}{N}\sum_{\bm{k},i\omega_{n}}\left[G_{11}(\bm{k},i\omega_{n})G_{33}(\bm{k}+\bm{q},i\omega_{n}+i\epsilon_{n}) +( \bm{k}\rightarrow \bm{k}-\bm{q})\right]\\
\nonumber
\Sigma_{2}(\omega,\bm{q}) & = \frac{g^{2}}{N}\sum_{\bm{k},i\omega_{n}}\left[G_{12}(\bm{k},i\omega_{n})G_{34}(\bm{k}+\bm{q},i\omega_{n}+i\epsilon_{n}) +(\bm{k}\rightarrow \bm{k}-\bm{q})\right]\\
\nonumber
\Sigma_{3}(\omega,\bm{q}) & =\frac{g^{2}}{N} \sum_{\bm{k},i\omega_{n}}\left[G_{13}(\bm{k},i\omega_{n})G_{31}(\bm{k}+\bm{q},i\omega_{n}+i\epsilon_{n}) +( \bm{k}\rightarrow \bm{k}-\bm{q})\right]\\
\nonumber
\Sigma_{4}(\omega,\bm{q}) & =\frac{g^{2}}{N} \sum_{\bm{k},i\omega_{n}}\left[G_{14}(\bm{k},i\omega_{n})G_{32}(\bm{k}+\bm{q},i\omega_{n}+i\epsilon_{n}) +( \bm{k}\rightarrow \bm{k}-\bm{q})\right],
\end{align}
 where $G_{ab}$ with $[a,b=1,2,3,4]$ are the various matrix elements of the Green's function matrix defined in Equation \eqref{eq_greenmat}.

The renormalized phonon propagators can be written using the Dyson equations obtained by using the above self-energies (Equation \eqref{eq:suple_pi}). 
The~Dyson equations are as follows~\cite{ref-PALeeSSC1993} (also see Figure~\ref{fig:Dyson}):

\begin{figure}[h]
\includegraphics[width=1\linewidth]{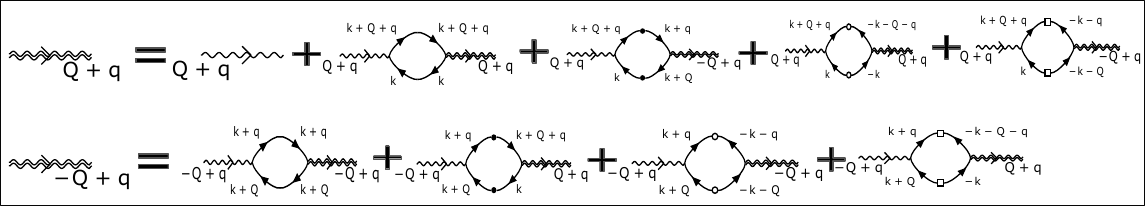}
\caption{Diagrammatic representations of the Dyson equations for phonons in the presence of charge density wave (CDW) and superconductivity (SC). The~double wavy lines represent the renormalized phonon modes. Each bubble diagram corresponds to the self-energy corrections $\Sigma_1$, $\Sigma_3$, $\Sigma_2$ and $\Sigma_4$, respectively, and their analytic expressions are given in Equation \eqref{eq:suple_pi}.}
\label{fig:Dyson}
\end{figure}

\text{~}\\
\vspace{-18pt}
\begin{align}\label{eq:Dyson}
\tiny
D_{1}(z,\bm{q}) &= D_{0}(z,\bm{q}+\bm{Q})\bigg[1 + \Sigma_{1}(z,q)D_{1}(z,\bm{q})+\Sigma_{2}(z,\bm{q})D_{1}(z,\bm{q}) \\
\nonumber
&+\Sigma_{3}(z,\bm{q})D_{2}(z,\bm{q})+\Sigma_{4}(z,\bm{q})D_{2}(z,\bm{q})\bigg],\\
\nonumber
 D_{2}(z,\bm{q}) & =  D_{0}(z,\bm{q}-\bm{Q})\bigg[\Sigma_{1}(z,\bm{q})D_{2}(z,\bm{q})+\Sigma_{2}(z,\bm{q})D_{2}(z,\bm{q}) \\
\nonumber
&+\Sigma_{3}(z,\bm{q})D_{1}(z,\bm{q})+ \Sigma_{4}(z,\bm{q})D_{1}(z,\bm{q})\bigg],
\end{align}
where $D_0(z,\bm{q}) = 2 \omega_{\bm{q}}/(z^2 - \omega_{\bm{q}}^2)$ is the bare phonon propagator. We obtain the renormalized modes for phonons in the presence of CDW and SC orders by~decoupling Equation~\eqref{eq:Dyson} through $D_{\pm}(z,\bm{q}) = D_{1}(z,\bm{q}) \pm D_{2}(z,\bm{q})$ and then solving $D_{\pm}(z,\bm{q})$ with the assumption that $\omega_{\bm{Q}\pm \bm{q}} \approx \omega_{\bm{Q}}$ for small $\bm{q}$. Finally, plugging in $D_{0}(z,\bm{q})$, we obtain the solutions for the renormalized phonon propagators as
\begin{align}\label{Eq:Modephonon}
D_{\pm}(z,\bm{q}) = \frac{2\omega_{\bm{Q}}}{z^{2}-\omega_{\bm{Q}}^{2}-2\omega_{\bm{Q}}\Sigma_{\pm}(z,\bm{q})},
\end{align}
where $\Sigma_{+} = \Sigma_{1}+\Sigma_{2}+\Sigma_{3}+\Sigma_{4} $ and $\Sigma_{-} = \Sigma_{1}+\Sigma_{2}-\Sigma_{3}-\Sigma_{4}$.
The dispersion of the renormalized phonon modes corresponds to the values of $z$ for~which the denominator of Equation \eqref{Eq:Modephonon} vanishes. Subsequently, keeping only $\bm{q}$ dependence in $\Sigma$, the~frequency for each mode is given by
\begin{align}\label{eq:dispersion}
\Omega_{\pm}^{2}(\bm{q})=\omega_{\bm{Q}}^{2}+2\omega_{\bm{Q}}\Sigma_{\pm}(\bm{q}).
\end{align}

We notice, that in the ordered phase where CDW and SC have formed, there are two normal phonon modes as given by Equation \eqref{eq:dispersion}; however, we will only focus on the $\Omega_{-}$ mode~\cite{ref-SarkarPRR2021}. We also notice from  Equation \eqref{eq:dispersion} that the change in the phonon frequency of the $\Omega_{-}$ mode is depicted through the self-energy correction $\Sigma_{-}(q)$, keeping $\omega_{Q}$ constant. Therefore, the~softening of the phonon frequency of the $\Omega_{-}$ mode implies a suppression in the  $\Sigma_{-}(q)$. Hence, it is sufficient to plot the $\Sigma_{-}(q)\equiv \Sigma_{}(q)$ instead of the full $\Omega(q)$, where $\bm{q}$ measures the deviation from $\bm{Q} [\bm{q} =0]$.

\section{Doping-Induced Fermi Surface Evolution of Phonon Softening in~Cuprates}\label{sss333}

In the last section, we described the general formalism for calculating the renormalized phonon frequency in the presence of CDW and SC orders. Now, in~this section, we concentrate on the specific case for the hole-doped cuprates. After~decades of research on the~phenomenology of the CDW phase, its correlations with superconductivity and pseudogap phase are strongly debated, and there is a common belief from theoretical~\cite{ref-efetovNat2013,ref-Wangprb2014} and experimental~\cite{ref-dasilvaScience2014,ref-CominSc2014} studies that the CDW wave vector is incommensurate and supports a momentum space structure where the Fermi surface plays crucial roles in defining the CDW wave vector $\bm{Q}$.
Within~this picture~\cite{ref-efetovNat2013,ref-Wangprb2014}, the~CDW wave vector $\bm{Q}$ is measured by connecting the `hot-spots' ($\bm{k}$-points where the Antiferromagnetic Brillouin zone boundary intersects the Fermi surface) on the Fermi surface, as~indicated in Figures~\ref{Fig:PhononLSCO}a and \ref{Fig:phononBSCCO}a by green arrows. In~this work, we consider CDW wave vector $\bm{Q}$, connecting the hot-spots and also considering them to be bi-axial, following our previous work~\cite{ref-SarkarPRR2021} for our model cuprate systems. Next, following the self-consistent results of~\cite{ref-Chakraborty2019,ref-Grandadam2020}, we consider that the CDW gap $\chi_k$ is maximum ($\chi_{max}$) near the hot-spots and falls off exponentially away from the hot-spots. The~SC gap is chosen such that $\Delta_{k}$ gaps out the rest of the Fermi surface; however, it vanishes along the nodal direction $(\pi,\pi)$ of the Fermi surface dictated by a d-wave symmetric SC gap, given by $\Delta_{k} = (\Delta_{max}/2)[ \cos(k_x) - \cos(k_y)]$ \cite{ref-scalapinoPH1995}. Below,~we separately consider Fermi surfaces' evolutions of phonon frequencies for two tight-binding band structures which mimic Bi$_{2}$Sr$_2$CaCu$_2$O$_{8 +\delta}$ (BSCCO) and La$_{2-x}$Sr$_x$CuO$_4$ (LSCO) systems.

\begin{table}[h]
\caption{Table for tight-binding parameters (in eV units) used in Equation \eqref{eqn:LSCOband} and resulting in the six Fermi surfaces, as plotted with solid lines in Figure~\ref{Fig:PhononLSCO}a.\label{Table:dopLSCo}}


\begin{tabular}[t]{ |p{1cm}||p{1.5cm}|p{1.5cm}|p{1.5cm}|p{1.5cm}|p{1.5cm}|p{1.5cm}| }
\hline
\multicolumn{7}{|c|}{Tight-binding parameters} \\
\hline
\textbf{Case:} & $t_1$ & $t_2$ & $t_3$ & $t_4$ & $t_5$ & $\mu$\\
\hline
1 & -0.7823 & 0.0740 & -0.0587 & -0.1398 & -0.0174 & 0.0801\\
\hline
2 & -0.7823 & 0.0740 & -0.0487 & -0.1398 & -0.0074 & 0.080\\
\hline
3 & -0.7823 & 0.0740 & -0.0287 & -0.1398 & -0.0044 & 0.0795\\
\hline
4 & -0.7823 & 0.0740 & -0.0187 & -0.1398 & -0.0024 & 0.0793\\
\hline
5 & -0.7823 & 0.0740 & -0.0087 & -0.1398 & -0.0014 & 0.079\\
\hline
6 & -0.7823 & 0.0740 & -0.006 & -0.1398 & -0.00038 & 0.0789\\
\hline
\end{tabular}

\end{table}

\subsection{Case of La$_{2-x}$Sr$_x$CuO$_4$}

To capture the generic features of the Fermi surface evolution towards a topological transition in  La$_{2-x}$Sr$_x$CuO$_4$, we consider a six-parameter tight-binding model band structure~\cite{ref-DevOdd2021} given by
\begin{align}\label{eqn:LSCOband}
\nonumber
\xi_{k} &= t_{1}\frac{\cos(k_x) + \cos(k_y)}{2} + t_2\cos(k_x)\cos(k_y) + t_3\frac{\cos(2k_x) + \cos(2k_y)}{2} \\
& + t_4\frac{(\cos(2k_x)\cos(k_y) + \cos(2k_y)\cos(k_x))}{2} + t_5(\cos(2k_x)\cos(2k_y)) + \mu.
\end{align}

\begin{figure}[h]
\includegraphics[width=0.97\linewidth]{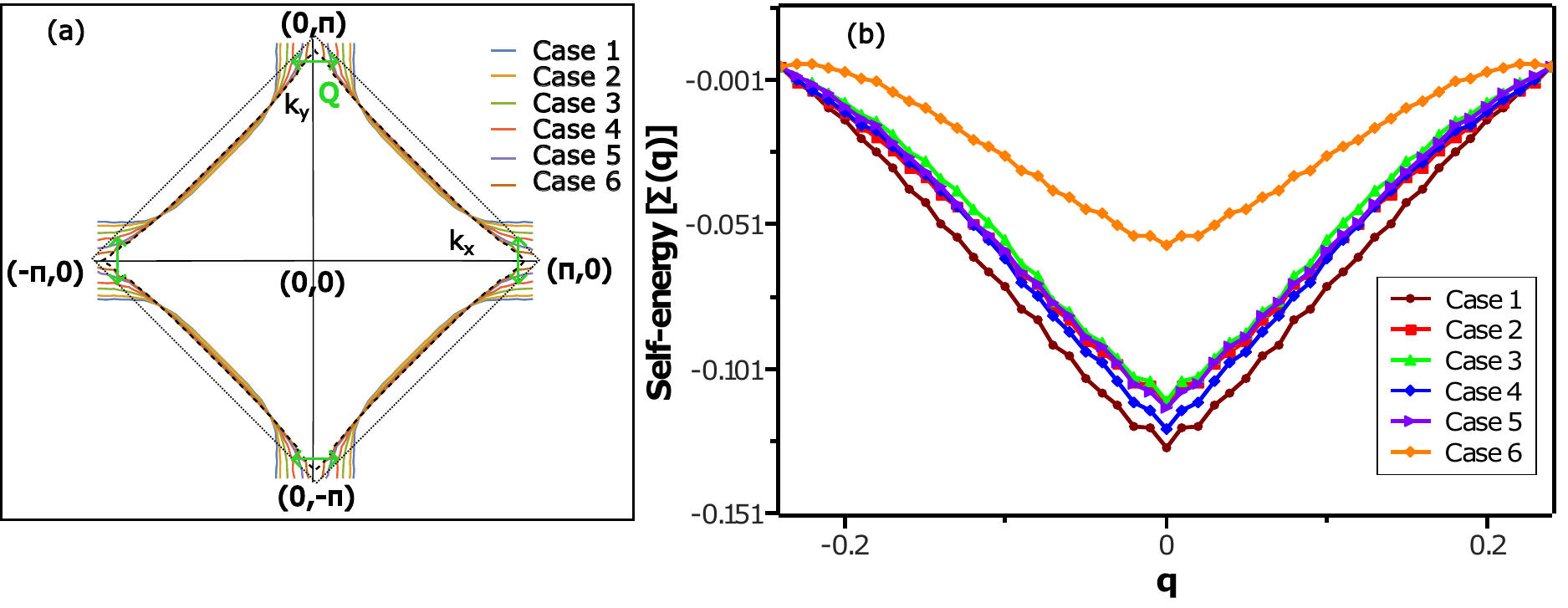}
\caption{(\textbf{a}) Fermi surfaces obtained from band structure Equation \eqref{eqn:LSCOband} with different tight-binding parameters as given in Table~\ref{Table:dopLSCo}. The~Fermi surface evolves from hole-like to electron-like with change in the parameters. The~electron-like Fermi surface is plotted with black dashed line. The~tight-binding parameters corresponding to the black dashed line are given in the main text. The~CDW wave vector ($\bm Q$) is indicated with the green arrow. (\textbf{b}) Phonon softening: plots for the evolution of the self-energy $\Sigma(q)$ as a function of $\bm{q}$ for the same sets of the tight-binding parameters used in (\textbf{a}) and given in
Table~\ref{Table:dopLSCo}. \label{Fig:PhononLSCO}}
\centering
\end{figure}

To mimic the doping evolution of the Fermi surface on a qualitative level, we perform minimal changes in the band structure (Equation \eqref{eqn:LSCOband}) by~changing only the hopping parameters $t_3$, $t_5$ and the chemical potential $\mu$. A similar evolution of the Fermi surface is seen in the ARPES experiments~\cite{ref-Fujimori1998,ref-YoshidaPRB2006} with a change in doping. The~tight-binding parameters for six different cases, replicating six different Fermi surfaces considered, are given in Table~\ref{Table:dopLSCo}. However, these tight-binding parameters are not fitted to the experimentally observed Fermi surfaces. The~resulting Fermi surfaces for these six cases are shown in Figure~\ref{Fig:PhononLSCO}a. We notice in Figure~\ref{Fig:PhononLSCO}a that the~Fermi surface for case 1 is hole-like featuring Fermi arcs. The~geometry of the Fermi surface changes from hole-like to electron-like, going from case 1 to case 6. We depict the closed electron-like Fermi surface with a dashed black line in Figure~\ref{Fig:PhononLSCO}a. The~tight-binding parameters corresponding to the dashed black line are $t_1 =-0.7823$, $t_{2}=0.0740$, $t_3 = -0.00387$, $t_4 = -0.1398$, $t_5 = -0.00037$ and $\mu = 0.0788$. To~compute the renormalized phonon frequency for the above six cases in Table~\ref{Table:dopLSCo}, we choose $\chi_{max}$ and $\Delta_{max}$ to be equal to 0.1 eV, a~realistic~value.  

 The computed phonon self-energies $\Sigma(q)$ as a function of $\bm{q}$ are plotted in Figure~\ref{Fig:PhononLSCO}b. First of all, we observe that $\Sigma(q)$ decreases strongly around $\bm{q}=0$, implying a phonon softening around $\bm{Q}$. Also, with~going away from $\bm{q}=0$, suppression in $\Sigma(q)$ is diminished, suggesting that phonon softening is reduced away from $\bm{Q}$. Next, we closely look at the evolution of the $\Sigma(q)$ from case 1 to case 5. We observe that the~strength of phonon softening remains more or less equal, as there are not many changes in  $\Sigma(q)$. However, if~we notice the Fermi surfaces corresponding to cases 1 to 5 in Figure~\ref{Fig:PhononLSCO}a, we observe that the hole-like Fermi surface continuously changes towards an electronic Fermi surface. Very interestingly, we notice that for case 6, where the Fermi surface (Figure~\ref{Fig:PhononLSCO}a) is in close vicinity to the topological Lifshitz transition, the~strength of the phonon softening (Figure~\ref{Fig:PhononLSCO}b) is significantly suppressed as compared to the earlier five cases. Hence, in~the case for the model LSCO band, the~strength of phonon softening changes discontinuously, remaining initially unaffected by the change in the geometry of the Fermi surface and then suddenly diminishing when the Fermi surface is very close to the Lifshitz transition~.
 
\vspace{-6pt}
 
 \subsection{ Case of Bi$_{2}$Sr$_2$CaCu$_2$O$_{8 +\delta}$}

\begin{figure}[h]
\hspace{-12pt}\includegraphics[width=0.96\linewidth]{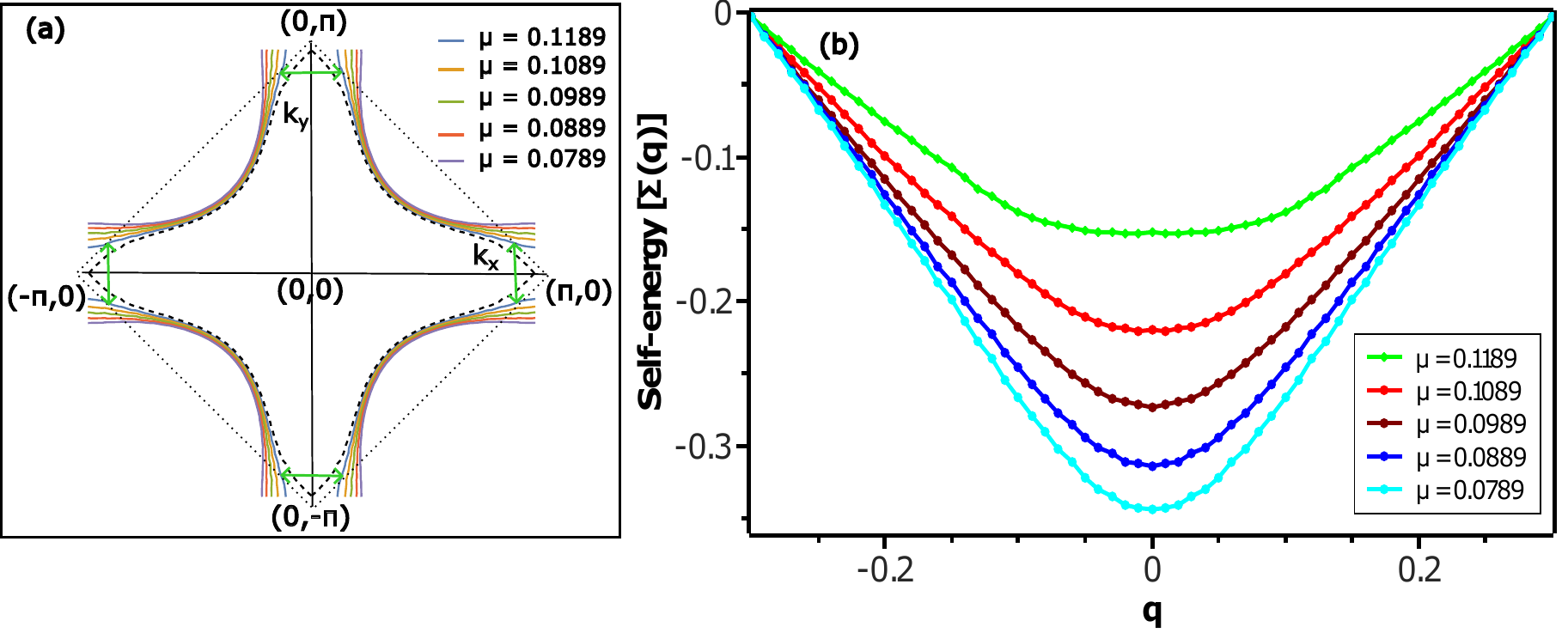}
\centering
\caption{(\textbf{a}) Tight-binding Fermi surfaces obtained from Equation \eqref{eqn:BSCCOband} for different values of $\mu$. A~continuous evolution from hole-like to electron-like Fermi surface can be seen with increase in $\mu$. Electron-like Fermi surface is shown in black dashed line for $\mu = 0.1329$. The~CDW wave vector ($\bm Q$) is indicated with the green arrow. (\textbf{b}) Phonon softening: plots of the self-energy $\Sigma(q)$ of phonons as a function of $\bm{q}$ and for the same values of $\mu$ used in (\textbf{a}).\label{Fig:phononBSCCO}}
\end{figure}

In order to capture the broad features of the evolution of the Fermi surface from hole-like to  electron-like for Bi$_{2}$Sr$_2$CaCu$_2$O$_{8 +\delta}$,  we consider a six-parameter tight-binding model band structure~\cite{ref-EschrigPRB12003} given by
\begin{align}\label{eqn:BSCCOband}
\nonumber
\xi_{k} &= t_{1}\frac{\cos(k_x) + \cos(k_y)}{2} + t_2\cos(k_x)\cos(k_y) + t_3\frac{\cos(2k_x) + \cos(2k_y)}{2} \\
& + t_4\frac{(\cos(2k_x)\cos(k_y) + \cos(2k_y)\cos(k_x))}{2} + t_5(\cos(2k_x)\cos(2k_y)) + \mu,
\end{align}
with $t_1 = -0.5908$, $t_2 =0.0962$, $t_3 = -0.1306$, $t_4 = -0.0507$ and $t_5 = 0.0939$. Here, for simplicity, we only change the chemical potential $\mu$ to capture the qualitative behavior of the topological transition of the Fermi surface, as observed in the ARPES experiment~\cite{ref-KAMPRB2006}; thus, the model band is not expected to exactly fit the ARPES bands~\cite{ref-KAMPRB2006}. The~tight-binding parameters for the band are in eV units. The~resulting Fermi surfaces for different $\mu$ values are displayed in Figure~\ref{Fig:phononBSCCO}a. We notice that the geometry of the Fermi surface continuously evolves from hole-like  to electron-like with increasing $\mu$ continuously from $0.0789$ to $0.1189$. On~further increasing $\mu$, the~Fermi surface goes through a topological Lifshitz transition. The~closed electron-like Fermi surface  corresponding to a $\mu$ value of $0.1329$ is plotted with a dashed black line in Figure~\ref{Fig:phononBSCCO}a.

To study the renormalized phonon frequency, we choose both $\Delta_{max}$ and $\chi_{max}$ to be 0.02 eV, a~value realistically obtained. In~Figure~\ref{Fig:phononBSCCO}b, we plot the self-energy $\Sigma(q)$ as a function of $\bm{q}$ for each of the $\mu$ values in Figure~\ref{Fig:phononBSCCO}a. We notice that for a $\mu$ value of $0.0789$, the~value of the $\Sigma(q)$ is suppressed around $\bm{q}=0$, and~the suppression is strongest at $\bm{q}=0$, that is, at the CDW wave vector. This implies that there is a phonon softening around the CDW wave vector. Once we go away from $\bm{q}=0$, the~phonon softening is significantly reduced as the suppression in $\Sigma(q)$ is diminished. Next, we see that with~an increase in the value of $\mu$, the~phonon softening becomes diminished continuously as the suppression in $\Sigma(q)$ around $\bm{q}=0$ becomes smaller with an increase in $\mu$ values. This implies that for the case of the model BSCCO band, the~phonon softening continuously diminishes as the hole-like Fermi surface evolves continuously towards the topological Lifshitz transition. This continuous evolution is in stark contrast to the case for the model LSCO band, where the phonon softening changes discontinuously as the hole-like Fermi surface evolves towards the topological~transition.

\section{Discussion}\label{sss444}
In this work, applying a perturbative approach, we study the Fermi surface evolution of the phonon softening in model cuprate bands, specially focusing on the parameter regimes where the Fermi surface approaches a topological Lifshitz transition. In~particular, we focus on two cuprate systems LSCO and BSCCO, where ARPES experiments also observed a topological transition from a hole-like Fermi surface to an electron-like Fermi surface due to a change in~doping. 

\begin{figure}[h]
\includegraphics[width=0.96\linewidth]{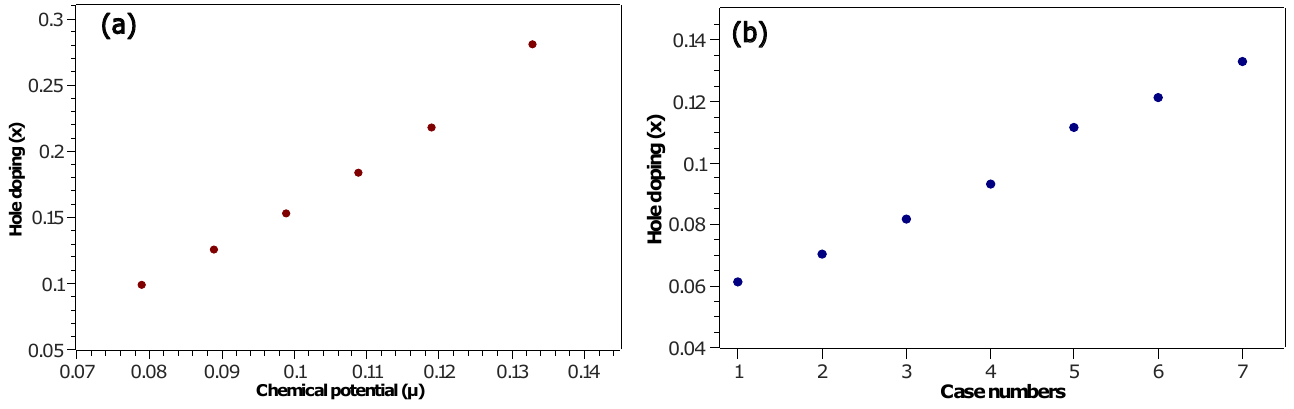}
\caption{(\textbf{a}) Plot of the hole doping ($x$) versus chemical potential ($\mu$) for the model tight-binding band in Equation  \eqref{eqn:BSCCOband} imitating BSCCO.
The~last point represents the doping around the topological Lifshitz transition in the theoretical model. (\textbf{b}) Plot of the hole doping ({$x$}) versus tight-binding parameters (shown in Table~\ref{Table:dopLSCo}) for the model band in Equation \eqref{eqn:LSCOband} imitating LSCO. The~last point represents the doping around the topological Lifshitz transition in the theoretical~model.\label{Fig:Doping_TB}}
\end{figure}

In this work, to~only focus on the possible effects of the change in the geometry of the Fermi surface due to doping on phonon softening, we have excluded the doping dependencies of the CDW and SC gaps. Furthermore, as~the theoretical tight-binding parameters are not fitted to experimental Fermi surfaces, the~doping levels ({$x$}) (as shown in Figure~\ref{Fig:Doping_TB}a,b) and number of holes ($p = 1+x$) per Cu atom associated with the parameters chosen in this work might differ from the experimental doping values where phonon softening~occurs.

For the case of the model system representing BSCCO, {\color{black}we} have found that the phonon softening {\color{black}is continuously suppressed} (Figure~\ref{Fig:phononBSCCO}b) with the continuous evolution of the Fermi surface from a hole-like to an electron-like geometry, and~it is significantly suppressed close to the topological transition. For~the model system of LSCO, we have found two features in the evolution of the phonon softening. First, the~phonon softening remains initially unaffected, even when the Fermi surface continuously changes from a hole-like towards an electron-like geometry, which is in sharp contrast to the case of BSCCO.
Second, only close to the topological transition of the Fermi surface, the~phonon softening becomes significantly suppressed. This difference in the nature of evolution of the phonon softening between these two materials can be mainly attributed to the difference in the band structure and associated differences in the nesting properties of the Fermi surfaces in these~systems. 

It would also be interesting to compare the results obtained from the theoretical models to the experiments. However, these comparisons must be considered at a qualitative level. A~recent RIXS experiment~\cite{ref-LuPRB2022} measured phonon softening in BSCCO with a change in doping at T$_c$ and below T$_c$. They found a gradual suppression of the phonon softening with an increase in doping from an under-doped to an over-doped regime at T$_c$. At~a low temperature, the~experiment again found a gradual suppression of phonon softening with an increase in doping in a broad window between the under-doped and over-doped regime, with~one exception in the highly under-doped regime. From~the calculations performed in this work for BSCCO bands, we have found a gradual suppression of phonon softening with an increase in doping, which is in general agreement to the above experimental findings. Noticeably, we have not found any enhancement of phonon softening in the under-doped regime within the parameter regime explored in this work. Therefore, it would be interesting to investigate whether incorporating a more accurate description of the band and doping dependence of the CDW and the SC orders and also the effects of CDFs can result in such an enhancement, as~seen in the extreme under-doped case in the~experiment. 

Another recent RIXS experiment~\cite{ref-LinPRL2020} measured the evolution of the phonon softening in LSCO in a broad doping regime across the topological Lifshitz transition. The~experimental results suggest that the strength of the phonon softening associated with the CDW remains almost the same with a gradual increase in the doping, even in the vicinity of the Lifshitz transition, but~it abruptly vanishes at a critical doping, where eventually the CDW order disappears. By~taking into account a theoretical model of a tight-binding band for LSCO and coupling between phonons with bi-axial CDW and SC, we have found that the strength of phonon softening remains approximately the same while the Fermi surface evolves continuously towards the Lifshitz transition. This bears a resemblance to the experimental observations. Moreover, we have also found that close to the topological Lifshitz transition, the~phonon softening {\color{black}is abruptly suppressed}. This feature is in contrast to the experimental~observations. 

In the context of LSCO, we now discuss some aspects which need further investigation. The~RIXS experiment~\cite{ref-LinPRL2020} in LSCO suggests a CDW-driven phonon softening, implying that in the doping regime considered in the experiment, there might be no signature of the magnetic order and only CDW correlations survive, which is also in accordance with previous experimental observations~\cite{ref-WenNat2019,ref-vonNPJ2023}. Hence, our theoretical model which involves coupling between phonons, CDW and SC is a consistent model for describing phonon softening in the vicinity of the topological transition in this material. However, the~effects of uni-axial stripy CDW~\cite{ref-EmeryPNAS1999,ref-TranquadaNat1995} pertinent to LSCO, which are not currently considered in our work, can play crucial roles in describing the differences. Additionally, more accurate descriptions of the doping dependence of bands and the~CDW and the SC orders can play unique roles. It should also be noted that while for LSCO, the~experimental~\cite{ref-LinPRL2020} wave vector corresponding to the strongest phonon softening slightly differs from the CDW wave vector ($\bm Q$), within~our theoretical framework, the~strongest phonon softening always occurs at the CDW wave vector $\bm Q$. This can be related to the fact that for simplicity we have considered in our model an~electron--phonon coupling constant which is independent of both fermion and phonon momenta, while for the breathing modes in cuprates, the~electron--phonon coupling can in general depend on the momenta~\cite{ref-DevreauxPRL2004} and can result in such an offset of~the wave vector.
 
Besides the doping dependence, another vital aspect of phonon softening in cuprates is their temperature evolution. The~RIXS experiment in BSCCO~\cite{ref-LuPRB2022} observed an enhancement in the phonon softening between T$_c$ and low temperature for some doping regime. The~RIXS experiment~\cite{ref-LinPRL2020} in LSCO measured phonon softening at T$_c$ and above T$_c$. Curiously, for~both BSCCO~\cite{ref-LeeNAT2021} and LSCO~\cite{ref-LinPRL2020}, the~phonon softening still survives above T$_c$. To~theoretically account for these behaviors at T$_c$ and above, one needs to include the effects of thermal fluctuations, the~short-range nature of the CDW and also the possible coupling between CDFs and the lattice, which all can play important roles. In~an earlier work~\cite{ref-SarkarPRR2021}, it was indeed found that fluctuation-driven damping effects and the interplay between the temperature dependence of CDW, SC and the damping can capture some crucial aspects of the temperature evolution of the phonon softening in the under-doped cuprates. As~in this work, we work at the mean-field level, and the~effects of higher temperatures cannot be properly accounted for and can also be partially responsible for the differences. Therefore, it will be an interesting future direction to theoretically study the combined temperature and doping evolution of the phonon softening in both BSCCO and~LSCO.

\section{Conclusions}\label{sss555}

To summarize, in~this work, by employing a perturbative technique, we have theoretically studied the effect of Fermi surface evolution close to the topological Lifshitz transition on phonon softening for two simple model tight-binding band structures broadly capturing doped LSCO and doped BSCCO systems in the presence of a charge density wave and superconductivity. We have found interesting features of phonon softening in these two materials and have compared them with experimental~observations. 

\acknowledgments{This work was supported by the Office of Basic Energy Sciences, Material Sciences and Engineering Division, U. S. Department of Energy, under Contract No.~DE-SC0012704.}

\end{document}